\documentclass[a4paper]{article}
\usepackage{epsfig,amsfonts,amssymb,setspace,multicol,textcomp,xcolor,multirow}
\usepackage[T1]{fontenc}
\usepackage{amsmath}
\makeatletter
\renewcommand{\@oddhead}{\textit{} \hfil \textit{O.\,V.\,Agapitov, O.\,K.\,Cheremnykh}}
\renewcommand{\@evenfoot}{\hfil \thepage \hfil}
\renewcommand{\@oddfoot}{\hfil \thepage \hfil}

\makeatother

\renewenvironment{thebibliography}[1]{\begin{oldthebibliography}{#1}\setlength{\parskip}{0ex}\setlength{\itemsep}{0ex}}{\end{oldthebibliography}}
\newcommand{\myHeder}[2]{\raisebox{+4cm}[.2cm][.2cm]{\vbox{\hbox to\textwidth{ #2 \hfill}\hbox to\textwidth{ #1 \hfill}}}}
\newcommand{\myCopy}[2]{\raisebox{-20.2cm}[.2cm][.2cm]{\vbox{\hbox to\textwidth{ #2 \hfill}\hbox to\textwidth{ #1 \hfill}}}}

\begin{document}
\fontsize{11}{11} \selectfont
\title{\bf Magnetospheric ULF waves driven by external sources}
\author{\textsl{O.\,V.\,Agapitov$^{1,2}$\thanks{oleksiy.agapitov@cnrs-orleans.fr}, O.\,K.\,Cheremnykh$^{3}$}}
\date{\vspace*{-10ex}}
\maketitle
\myHeder{\hspace*{-5ex} \fontsize{8}{8}\selectfont }{}
\myCopy{\fontsize{8}{8}\selectfont\copyright\hspace*{1ex}\sl O.\,V.\,Agapitov, O.\,K.\,Cheremnykh, 2013}{}
\begin{center}
{\small $^{1}$LPC2E/CNRS, University of Orleans, France\\
$^{2}$Taras Shevchenko National University of Kyiv, Glushkova ave., 4, 03127, Kyiv, Ukraine \\
$^{3}$Space Research Institute NASU-NSAU, Glushkova ave. 32, Kyiv 03022, Ukraine\\}
\end{center}
\vspace*{-4ex}
\begin{abstract}
The multi-spacecraft missions (Cluster and THEMIS) observations allowed to collect large data base for Ultra Low Frequency (ULF) waves properties, their  localization, and sources. Mainly here we focused on these recent results. Studying of the source and characteristics of ULF waves can help in the understanding of the interaction and energy transport from solar wind to the magnetosphere. Here we present peculiarities of ULF waves generated by different solar wind phenomenon: surface magnetopause instability, magnetosphere cavity modes and solar wind dynamic pressure sudden impulses (SI) penetration into the magnetosphere. Permanent observations of ULF waves involve existence of the permanent source and, as the previous studies showed, the contributions to Pc4-Pc5 ULF wave power from the external sources are larger than the contribution from internal magnetosphere sources. The Kelvin-Helmholtz instability (KHI) can generate on the magnetosphere flanks classical ULF resonant waves with spatially
localized amplitude maximum. As observations show the constraint satisfaction of KHI development is quite rare. SI in solar wind dynamic pressure generate ULF waves with different polarization and frequency close to the frequency of the local field line resonance (FLR). Wide range of temporal and amplitude characteristics of solar wind dynamics can generate magnetosphere cavity modes and magnetosonic perturbations which penetrate through the magnetosphere and can couple with the local FLR modes. The observed dependence of ULF waves properties on their localization are well correspond to these sources and their occurrence.\\[1ex]
\end{abstract}
\begin{multicols}{2}

\section*{\sc introduction}
\vspace*{-1ex}
\indent \indent Ultra low frequency (ULF) pulsations in the period range from 1\,s to more than 600\,s are one of the MHD modes via
which the magnetosphere reacts to solar wind dynamics and associated instabilities of the magnetospheric system. Here we present short review on resent results concerning the solar wind sources of magnetosphere ULF waves (see also the reviews \cite{Takahashi_etalJGR1998} with results of papers before Cluster missions, \cite{Menk} with review of the results published before 2010, for more details of wave-particle interaction see \cite{Elkington,Shprits2008}). The waves interact with trapped particles and this interaction is one of the factors which controls the dynamics of energetic electrons in the outer radiation belt. Studying factors that control magnetospheric ULF waves can provide better understanding how the waves are excited and how the solar wind affects the behaviour of magnetospheric particles. Statistical and case studies show that periodic variations in the solar wind dynamic pressure can directly guide magnetospheric ULF waves, especially on CMS frequencies: 0.7, 1.4, 2.0\,mHz, which can be significantly detected in the day side magnetosphere \cite{Agapitov_etal2008,Agapitov_etal2006, Cheremnykh2004, Claudepierre_etal2008, Claudepierre_etal2010,  Mthembu_etalAG2009, Viall_etalJGR2009, VillantePiersantiJASTP2011, Zhang_etal2010}. These perturbations can be either coherent oscillation in the solar wind impinging on the magnetopause \cite{Agapitov2009, KepkoSpence2003, Mthembu_etalAG2009}. The strong relation of the Pc4-5 wave power to solar wind velocity also have been confirmed statistically \cite{Lui_etal2009,Pahud_etalJASTP2009}. Correlation and regression analysis show that several parameters can play a role in the ULF waves generation and amplification: interplanetary magnetic field Bz component \cite{Simms_etal2010} and solar wind density for Pc3 \cite{DeLauretis_etalJGR2010, Heiliget_alAG2010}.

One of the paradigms of ULF pulsation theory is the resonant coupling of a compressional surface wave with toroidal oscillations somewhere deeper in the magnetosphere. This field line resonance (FLR) mechanism was first suggested in \cite{Tamao1965} and later used in \cite{Southwood1974} and \cite{ChenHasegawa1974} to interpret observational results \cite{Samsonetal1971}. Recent analytical studies in \cite{Cheremnykh2010,CheremnykhAgapitov2012,Leonovich2001} showed that in an dipole magnetic geometry an exact solution of small transverse and large longitudinal scales of perturbations  can exist, which describes transversely small-scale perturbations with the displacement vector lying in a magnetic surface. It is established that in a dipole magnetic field the perturbations with just two polarizations can exist. For poloidal polarization the equations of small oscillations are obtained coupled with magnetosonic component \cite{Cheremnykh2010,CheremnykhAgapitov2012,Cheremnykh2006a,Klimushkin_etal1998,
Klimushkin_etal2008,Leonovich2001a,Verkhoglyadova_etal1999}.

\begin{figure}[!h]
\centering
\begin{minipage}[t]{.99\linewidth}
\centering
\epsfig{file = 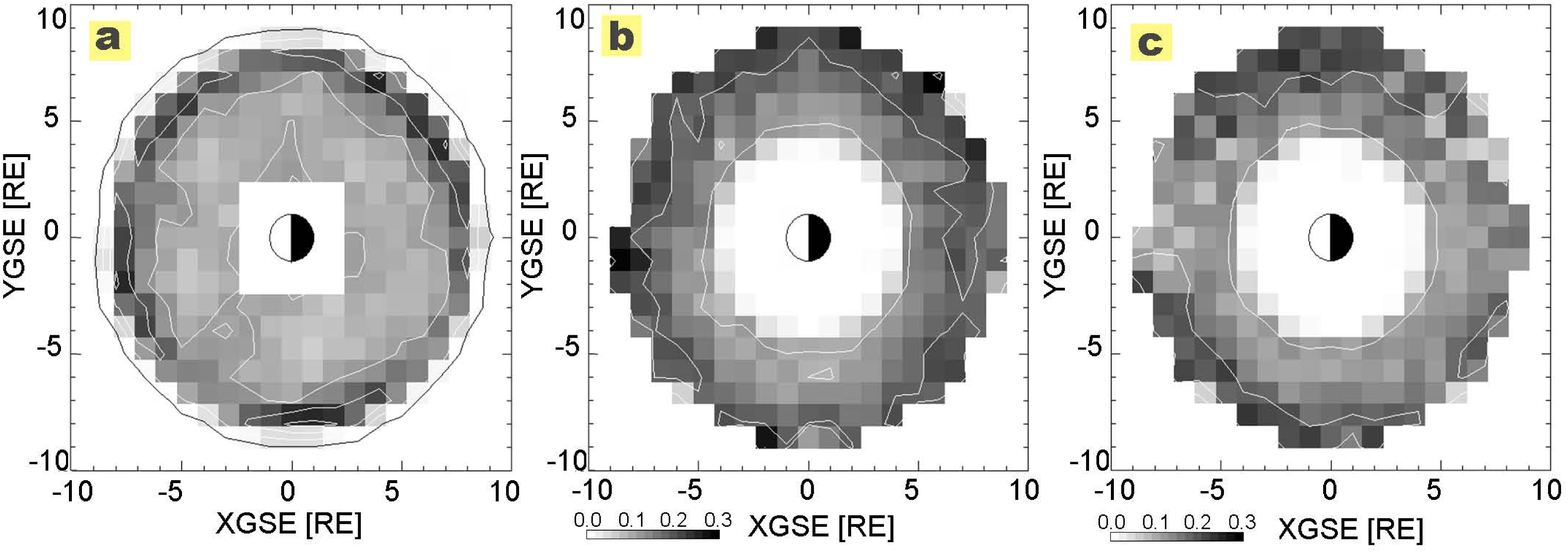,width = .99\linewidth}
\caption{Statistical results of the occurrence rate of Pc5 waves from AMPTE/CCE measurements: $a$ -- coverage of AMPTE/CCE measurements; $b$ -- occurrence rate of poloidal Pc5 waves; $c$ -- occurrence rate of toroidal Pc5 waves. }\label{ULFDistr}
\end{minipage}
\end{figure}

At the ground and in the ionosphere field line resonance associated oscillations are usually identified as latitudinally localized oscillations of the H-component of the magnetic field \cite{Green1982, Mathie_etal1999a, Rae_etal2005,Samsonetal1971} or in the NS-component of the ionospheric electric field \cite{Fenrich_etal1995,Walker_etal1979}, associated with a $180^{\circ}$ phase shift across this resonantly oscillating field line shell. In space, however, direct observations of field line resonances are very sparse \cite{Agapitov_etal2009, Lui_etal2009}. As direct observations we define spatially and temporally resolved observations of wave fields exhibiting typical spatial and temporal characteristics of field line resonances  \cite{Agapitov_etal2009}. A basic ingredient of the process is the resonant interaction of a compressional mode, spatially decaying towards the inner magnetosphere, and its coupling to a localized Alfv{\'e}nic perturbation at a point where the local field line shell eigenfrequency equals the frequency of the driving surface wave. The typical characteristics of a global FLR are predominantly toroidal magnetic field oscillations, localized in radial direction within the magnetosphere. Across the position of maximum field amplitude, the phase of the toroidal component changes by $180^\circ$ and the direction of polarization is reversed \cite{Nishida1978}. The spatially decaying surface mode is thought to be generated by the Kelvin-Helmholtz instability (KHI) of the dawn and dusk magnetopause \cite{Fujita1996,PuKivelson1983,Southwood1968,Walker1981}, conditions for which are best when passing through a high-speed solar wind stream \cite{Engebretson_etal1998, Pahud_etalJASTP2009,Seon_etal1995}. A ULF pulsation generated by the described resonant coupling between a compressional surface wave and a toroidally polarized and spatially confined Alfv{\'e}nic perturbation we define here as a classical field line resonance. Earlier studies, e.\,g. \cite{Singer1982}, reported radially localized, but poloidally polarized ULF waves using ISEE1 and ISEE2 observations and allowed to estimate the resonance region width to be about 0.7\,RE. \cite{Cramm_etal2000} also reported such localized poloidal modes close to the plasmapause and were able to demonstrate a significant phase variation across such a localized ULF wave event. But these events do not represent classical field line resonances, as the observed poloidal polarization was not in agreement with the theoretically expected toroidal polarization \cite{Southwood1974}. However, such spatially localized and poloidally polarized waves are related to spatial variations of the Alfv{\'e}n wave velocity close to the plasmapause \cite{Klimushkin_etal2004, Schaefer_etal2007} or can be excited by ring current ions on the dusk side of the magnetosphere \cite{Hudson_etal2004}. Further very convincing evidence for resonant toroidal oscillations in the dawn and dusk magnetosphere sectors was provided by \cite{Engebretson_etal1986}, who presented harmonically structured toroidal ULF waves with frequencies changing with $L$-shell, much as expected for classical field line resonances. AMPTE/CCE had full coverage of magnetic local time in the vicinity of geomagnetic equator up to $L$-shell equal to 8.8\,$R_E$. The distribution of Pc5 ULF waves in the magnetosphere obtained from AMPTE/CCE measurements is shown in Figure~\ref{ULFDistr}. In the flank regions the Pc5 occurrence rate is dominated by the toroidal mode, and in the noon sector Pc5 ULF wave occurrence rate is dominated with poloidal mode. Similar peculiarities were obtained from  CRRES measurements in \cite{Hudson_etal2004} and from THEMIS measurements in \cite{Lui_etal2009}. Such a distribution can be explained in terms of different generation
sources for different regions.

\vspace*{-1ex}
\section*{\sc ULF waves generation by KHI}
\begin{figure}[!h]
\centering
\begin{minipage}[t]{.99\linewidth}
\centering \epsfig{file = 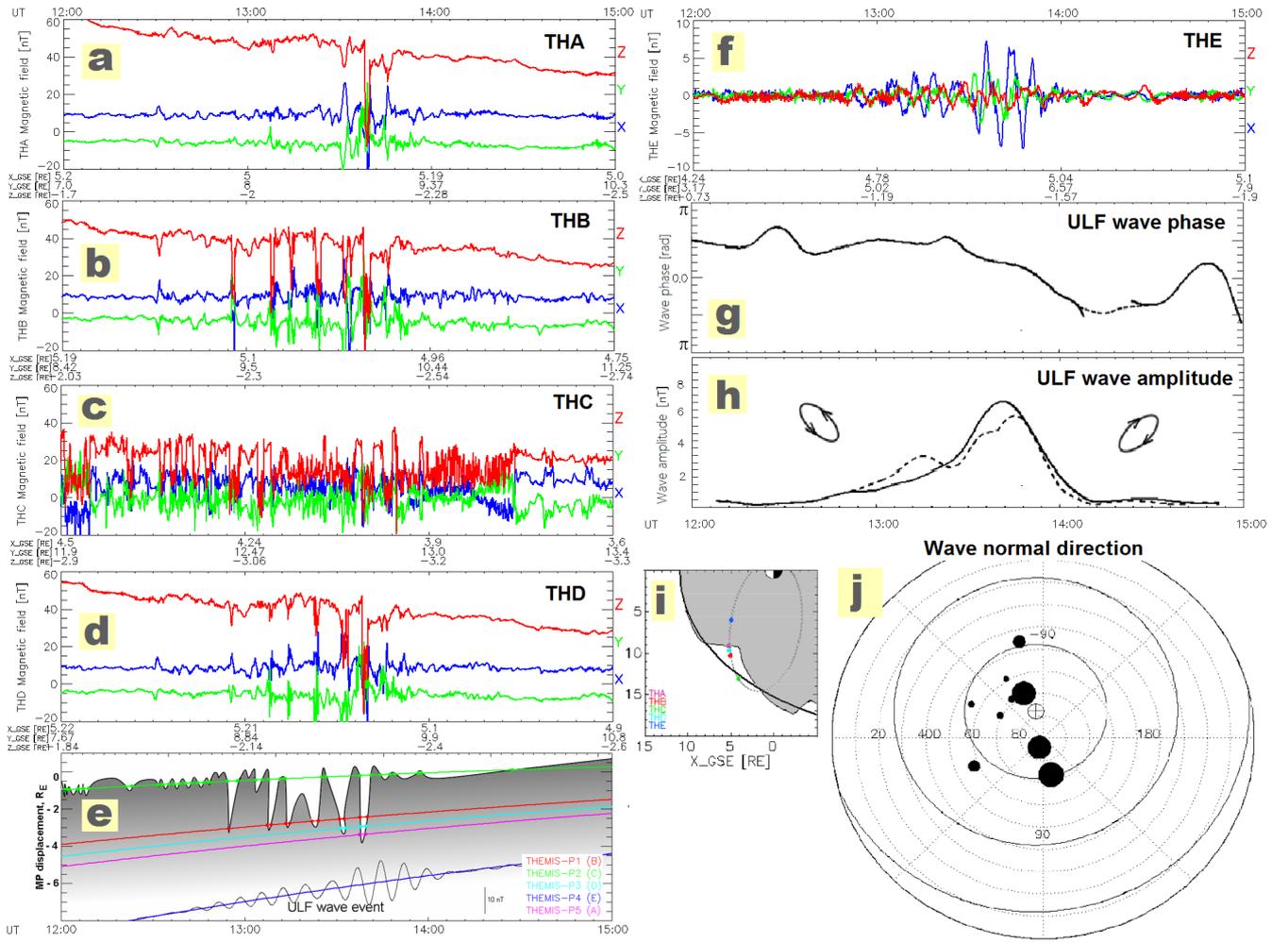,width = .99\linewidth}
\caption{Magnetic field measurements GSE components made aboard the THEMIS spacecraft THA ($a$), THB ($b$), THC ($c$), and THD ($d$) during the time interval on April 25, 2007, from 12:00-15:00~UT near the magnetopause.  The critical velocity according for Kelvin-Helmholtz instability development obtained from magnetosheath and magnetosphere conditions is 270-280\,km/s. The value of the magnetosheath plasma flow from THEMIS measurements is 290-310\,km/s. A schematic reconstruction of the magnetopause position using the observed magnetopause crossings and the Fairfield-71 magnetopause model is shown in panel $e$. Radial Positions of the different s/c are given with respect to the model magnetopause used. The magnetospheric wave event as seen by THE is also shown along the THE track. The THE magnetic field perturbations are shown in panel
$f$ in the field aligned coordinate system (red -- along the background magnetic field, blue -- toroidal perturbations). Analytical signal analysis results of THE magnetic field measurements: phase difference between the ${\rm B_{X}}$ and ${\rm B_{Y}}$ components in the field aligned coordinate system ($g$) and amplitude of the transverse magnetic field component for the frequency range $1.8-2.1$\,mHz ($h$). The polarization is schematically shown as well. Spacecraft trajectories during the observed time interval are shown with dashed lines in panel $i$. Magnetopause local boundary plane observed during spacecraft crossings is shown. The magnetopause crossing time moments are listed. The schematic recon
struction of the MP surface perturbation is shown on the top panel. Panel $j$ shows the directions of the $\mathbf{k}$ vectors of the ULF wave observed during 13:00-14:00\,UT, April 25, 2007. The Sun is to the right; the polar axis is in the ${\rm Z_{GSE}}$ direction. The background magnetic field direction is shown with the cross in the circle. The 30$^{\circ}$, 60$^{\circ}$, and 90$^{\circ}$ angles from the background magnetic field are shown with the solid lines. The comparative wave power is shown with the circle radius}\label{KHI}
\end{minipage}
\end{figure}

\vspace*{-1ex}
\indent\indent The examples of a KHI driven resonant ULF pulsation in space is the long-lasting event, reported about and analyzed in detail in \cite{Agapitov_etal2009,Rae_etal2005}. Multi-point observations from the CLUSTER and POLAR spacecraft allow to demonstrate that KHI driven surface waves drive resonant field line oscillations. The data presented in this study provide evidence for an observational link between wave activity at the magnetopause, in the magnetosphere, and in the ionosphere. The ULF event observed by THEMIS spacecraft system could be identified so far which shows all the characteristics expected for a classical field line resonance generated by the KHI rising on the magnetopause surface (Figure~\ref{KHI}). This event has been analyzed in detail in \cite{Agapitov_etal2009}, next to a ULF event analyzed in \cite{Rae_etal2005}, is probably the only observed example of a field line resonance (Figure~\ref{KHI}) in space most likely generated directly by clearly identified surface wave (Figure~\ref{KHI}a,b,c,d). The April 25, 2008 event certainly qualifies to directly demonstrate the functionality of the classical field line resonance process. THEMIS multi-spacecraft observations made in the time interval April -- September 2007 provided for a unique opportunity to reconstruct the magnetopause surface dynamics (see also \cite{Agapitov2009, Plaschke_etal2008, Voshchepynets}). During this time interval the THEMIS spacecraft crossed magnetopause surface more than 300 times. About half of the crossings were multi-crossings (several inward and outward boundaries crossing of single spacecraft during short time interval). Multipoint spacecraft observations reveal permanent quasi-periodic surface perturbation at the magnetopause \cite{Agapitov2009}, and particularly at ``magic'' Pc5 frequencies \cite{Plaschke_etal2009}, show that KHI driven classical field line resonances did exist and highlight its role for ULF waves generation near the flanks at solar minimum \cite{Lui_etal2009}. Global MHD simulations in the realistic magnetosphere presented in \cite{Claudepierre_etal2008} confirm that at constant solar wind speed two coupled modes of KHI surface waves may be generated near the magnetopause flanks. Nevertheless, the many observations of FLRs at the ground and in the ionosphere on the one hand and the sparsity of space observations of classical field line resonances on the other hand is worthwhile to be noted. Here, we present observations from the five THEMIS spacecraft \cite{Angelopoulos2008}, traversing the dusk magnetosphere on a highly elliptic orbit with the spacecraft very often aligned in almost radial direction during the coast phase of the mission. Such radial conjugations are ideal for studying surface waves and associated field line resonances. The global ULF pulsations similar to \cite{Mann_etal2002} and \cite{Rae_etal2005} with detail analysis of the magnetopause surface perturbations is presented. The simultaneous analysis of magnetopause perturbation and ULF waves inside the magnetosphere gave an opportunity to explain the properties of observed ULF waves and distinguish the different possible generation mechanisms.

\vspace*{-2ex}
\section*{\sc ULF waves generation by SI}
\vspace*{-1ex}
\indent\indent The direct driving of the ULF activity by the solar wind quasi-periodic density perturbations is discussed in \cite{Erikkson2006,KepkoSpence2003,Viall_etalJGR2009}. The  statistical results based on 10 years of observation in both the solar wind and the magnetosphere confirm that, while discrete frequencies across the entire analyzed range from 0.5 to 5.0\,mHz occur, certain sets (0.7, 1.4, 2.0, and 4.8\,mHz in the solar wind and 1.0, 1.5, 1.9, 2.8, 3.3, and 4.4\,mHz in the magnetosphere) occur more often than others and in 54\% of the solar wind data segments in which a spectral peak was identified, at least one of the same discrete frequencies was statistically significant in the corresponding magnetospheric data segment. Thus the conclusion made in \cite{Viall_etalJGR2009} is that discrete magnetospheric oscillations in the range from 0.5 to 5.0\,mHz range are directly driven by periodic solar wind number density structures 54\% of the time that the solar wind contains periodic number density structures. In \cite{Plaschke_etal2009}  a statistical analysis of 452 THEMIS observations of magnetopause oscillations over 8 months has been carried out. The discrete frequencies found are close to solar wind set of frequencies and partially confirm the results of \cite{Viall_etalJGR2009}. Actually the quasi-periodic magnetopause oscillation (with ``wave-length'' in the same order of magnetosphere size) shows the first stage of coupling of solar wind density periodicity with magnetosphere ULF wave activity. The outstanding question is determining the mechanism that causes these periodic number density structures in the solar wind to occur at particular length scales \cite{Viall_etalJGR2008}.

\begin{figure}[!h]
\centering
\begin{minipage}[t]{.99\linewidth}
\centering \epsfig{file = 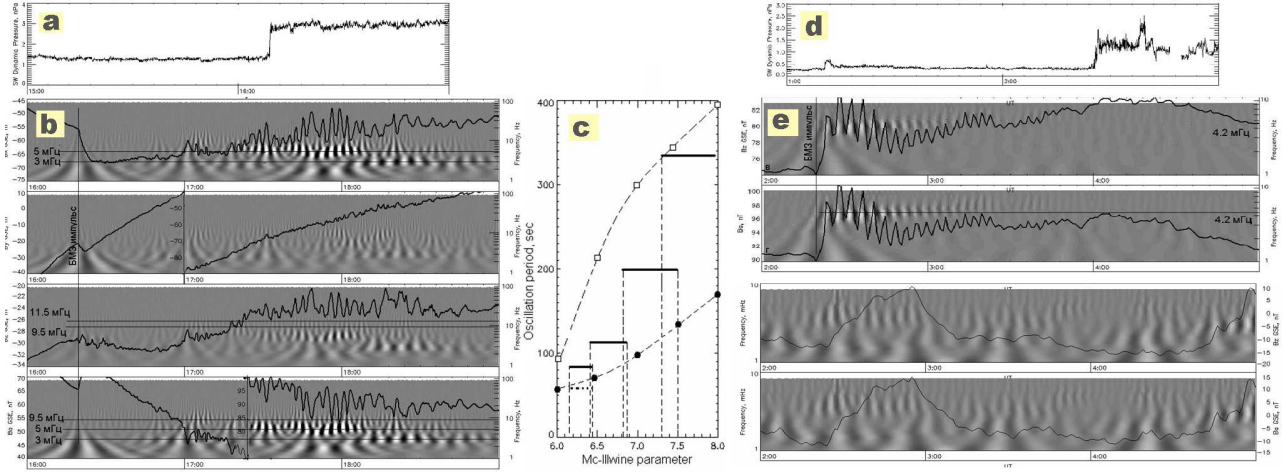,width = .99\linewidth}
\caption{SI in the solar wind dynamic pressure and their influence on ULF wave activity. $a$ -- SI in the solar wind dynamic pressure on detected aboard WIND spacecraft on January 10, 2001; $b$ -- magnetic field components and magnitude captured aboard Polar spacecraft on January 10, 2001; $c$ - dependence of FLR frequencies of toroidal (filled squares) and poloidal modes (white squares). The observed on January 10, 2001 aboard Polar spacecraft frequencies of ULF waves are shown by solid horizontal lines; $d$ -- SI in the solar wind dynamic pressure on detected aboard WIND spacecraft on January 13, 2001; $e$ -- magnetic field component along the background magnetic field and the magnetic field magnitude measured aboard GOES10 spacecraft on January 13, 2001; $f$ -- parallel component of the magnetic field and the magnetic field magnitude detected by ground based magnetometric measurements on Meanook near the magnetic conjugate point with GOES10 on January 13, 2001.}\label{SI_SW}
\end{minipage}
\end{figure}

In terms of natural oscillations in the magnetosphere their generation can be originated by effect of an external source with wideband spectrum. Fast changes of solar wind dynamic pressure can be considered to be a source with such characteristic. The increasing of solar wind dynamic pressure increases the surface currents on the magnetopause. As a result the magnetic field in the magnetosphere increases. In this case the magnetopause became a source of generation of several types of ULF waves. The pressure disturbance propagates through a magnetosphere with speed of fast MHD wave \cite{Nishida1978}. In the day sector of magnetosphere this velocity is close to Alfv\'{e}n speed (from 400 up to 10000\,km/s) and can leave behind shockwave in solar wind. Velocity of shock wave propagating in solar wind is usually in range 400-800\,km/s. However, sudden changes of the solar wind dynamic pressure, that is wideband perturbations, are discussed as an alternative to the KHI source mechanism \cite{
KivelsonSouthwood1985, SinhaRajaram2003}. Such perturbations can generate natural modes of the magnetospheric resonator with the polarization depending on the propagation direction and magnetic field disturbance vector \cite{AgapitovCheremnykh2008,Agapitov_etal2008,Agapitov_etal2008a, Klimushkin_etal2007,Mager_etal2006,Sarris}. ULF pulsations with different frequencies were observed simultaneously on different magnetic latitudes after a sudden impulse (Figure~\ref{SI_SW}) \cite{AgapitovCheremnykh2008,GlassmeierVolpers1984,Klimushkin_etal2004}. The existence of such spectral maxima affirms the magnetospheric property of selecting particular spectral peaks with global modes coupling to corresponding local field line shell oscillations \cite{AgapitovCheremnykh2008,Klimushkin_etal2012}. Qualitative confirmation of the experimental results were obtained extensive numerical modelling in a dipole magnetosphere \cite{LysakLee1997} and \cite{Allan_etal1986}. The recent numerical studies provided in \cite{Claudepierre_etal2008, Claudepierre_etal2010} showed the global structure of perturbation in the magnetosphere. The statistical study of magnetospheric effects of positive and negative solar wind pressure impulses was presented in \cite{Zhang_etal2010}. 270 ULF events excited by positive solar wind dynamic pressure pulses and 254 ULF events excited by negative pulses during 2001--2009 on geostationary orbit were analyzed and numerical simulation was provided. Both numerical and experimental results show efficiency of solar wind dynamic pressure pulses for ULF waves generations with toroidal and poloidal polarization, but magnetospheric response to positive pulses is much stronger that to negative ones. The poloidal waves are usually have larger amplitudes than toroidal. The amplitudes of ULF waves observed in the noon sector are larger than at dawn and dusk.

\vspace*{-2ex}
\section*{\sc ULF waves generation \\by waveguide modes}
\begin{figure}[!h]
\centering
\begin{minipage}[t]{.99\linewidth}
\centering \epsfig{file = 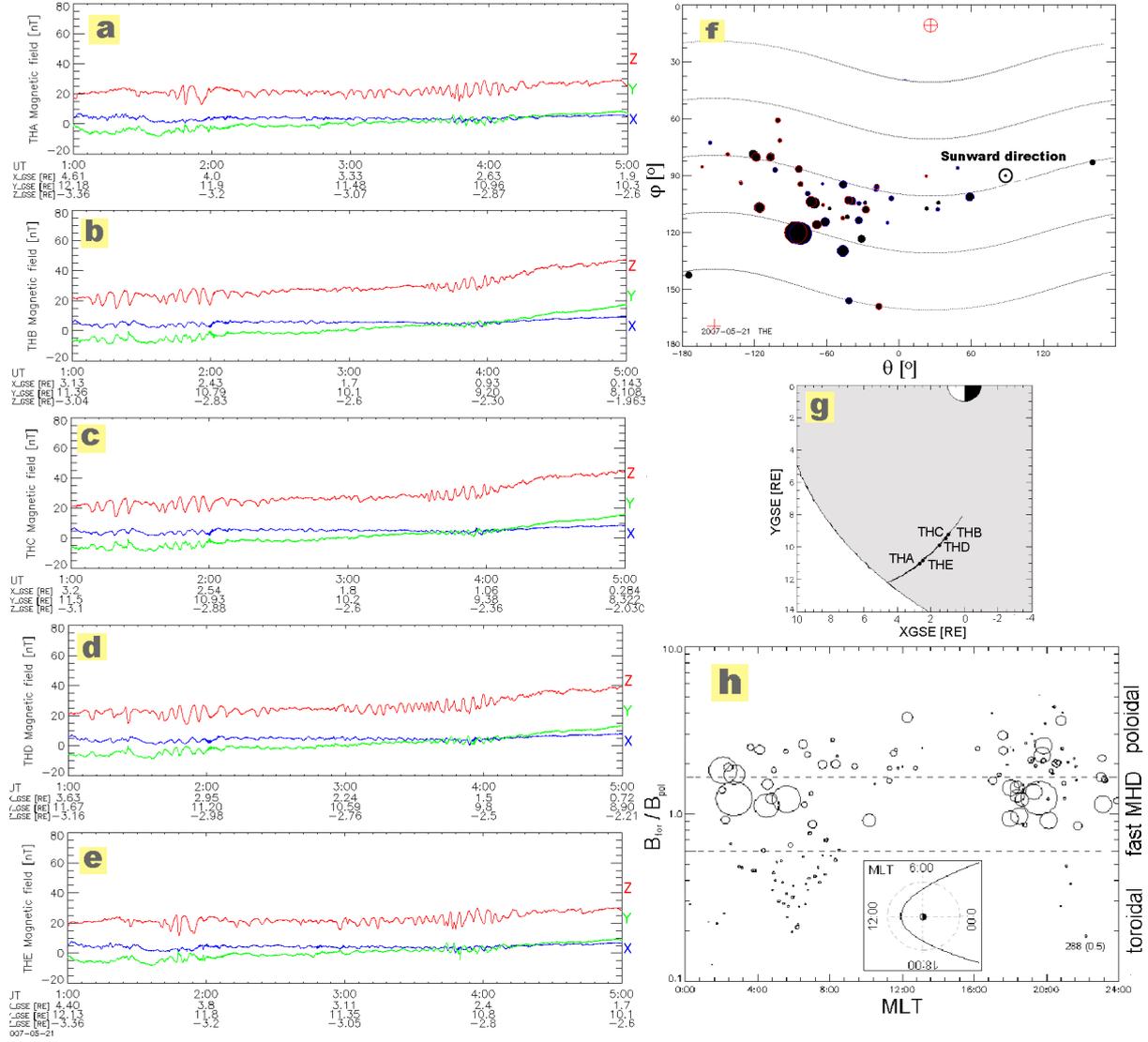,width = .99\linewidth}
\caption{Magnetic field measurements GSE components made aboard the THEMIS spacecraft THA ($a$), THB ($b$), THC ($c$), THD ($d$), and THE ($e$) during the time interval on May 21, 2007, from 1:00-5:00~UT on the magnetosphere flank region. The estimated directions of $\mathbf{k}$ are shown by filled circles (radius indicates wave amplitude) in the panel $f$. The direction of the background magnetic field is indicated by red cross in circle. Positions of the THEMIS spacecraft are shown in panel $g$. The statistics of the Pc5 ULF waves polarization in dependence on MLT is shown in panel $h$ as the ratio of poloidal component to toroidal one. The radius of circle shows the ratio of the parallel component of the magnetic field perturbation to the transverse one. }\label{WaveGuide}
\end{minipage}
\end{figure}

\vspace*{-1ex}
\indent\indent Further possibilities to drive discrete frequency perturbations with frequencies in the ULF range are cavity and waveguide modes \cite{Mann_etal1999, Mills_etal1999}. The waveguide theory predicts that the magnetosphere can act as a cavity which traps discrete frequency compressional mode energy between the magnetosphere boundary and the reflection region inside the magnetosphere \cite{Mann_etal1999}. Evidence of the magnetosphere cavity modes presented in \cite{Takahashi_etalJGR2010} shows the possibility to generate such eigen oscillations of the dayside magnetosphere just by non periodic changes of the solar wind dynamic pressure \cite{Claudepierre_etal2010,Voshchepynets}. Such trapped waves in the dayside magnetosphere can penetrate to the magnetosphere flank regions as waveguide magnetosonic modes. Most of the events analyzed by \cite{Mathie_etal1999b} using ground based observations can be explained in terms of such waveguide modes and agree closely with the theory proposed by \cite{Mann_etal1999} and \cite{Samson_etal1992}, having discrete frequencies of oscillation. CLUSTER measurements \cite{Mann_etal2002} furthermore support the hypothesis that, during intervals of fast solar wind speed, the KHI can excite magnetospheric waveguide modes which bathe the flank magnetosphere with discrete frequency ULF wave power and drive large amplitude resonant ULF pulsations. Multi-point observations show evidence of upstream waves entering and propagating through the magnetosphere as fast magnetosonic waves with different frequencies \cite{Constantinescu_etal2007AG,Hartinger2011,Voshchepynets}. The event detected by THEMIS spacecraft is shown in Figure~\ref{WaveGuide} ($a$-$g$). The quasi-periodic compressional magnetic field perturbations with frequency 3-4\,mHz were observed on the distances from 2 to 4~$R_E$ from the magnetopause  (Figure~\ref{WaveGuide}$g$). The estimated phase velocity was about 150-200\,km/s, that much less than the local Alvf{\'e}n velocity. Direction of propagation was anti-sunward. The observed frequencies are close to the FLR frequency range and coupling with resonant energy transport is possible.

\vspace*{-2ex}
\section*{\sc conclusions}
\vspace*{-1ex}
\indent\indent The progress in the understanding of the experimental properties of ULF waves during recent years on the basis of multi-spacecraft missions observations allows to distinguish the wave generation sources and now main questions are related to the efficiency of generation mechanisms (boundary instabilities, solar wind discontinuities, cavity modes, solar wind dynamic pressure oscillations) and solar wind conditions favorable for rather permanent ULF wave activity (see also \cite{Menk, Takahashi_etalJGR1998}). The statistical studies showed direct connection of solar wind dynamic pressure oscillations with ULF waves at discrete frequencies below 5\,mHz but their role in the everyday ULF wave activity and the nature of the such discrete frequencies presented in the solar wind is not clear now. The ULF occurrence rate and amplitude distributions obtained from different missions are rather similar: AMPTE/CCE, CRRES \cite{Hudson_etal2004}, THEMIS \cite{Lui_etal2009}, and well explained in terms of listed here generation mechanisms. Additional information can be obtained from polarization analysis (Figure~\ref{WaveGuide}$h$) \cite{AgapitovCheremnykh2011}: the ratio of poloidal component to toroidal one is shown in dependence on MLT. The radius of circle shows the ratio of the parallel component of the magnetic field perturbation to the transverse one. In the flank regions the wave power is dominated by the toroidal mode associated with KHI and SI. In the noon sector Pc5 ULF wave power is dominated with poloidal mode associated with the solar wind dynamic pressure SI and magnetosphere cavity modes. In the flanks also the fast MHD modes are observed propagating in mainly in the vicinity of the equator. Statistical correlation analysis indicates dependence of ULF waves amplitude and occurrence rate on solar wind parameters. These properties can be used to identify the sources of the observed modes obtained from single and multi-spacecraft measurements and further to develop the empirical model of ULF waves activity which is necessary for complex radiation belt dynamic model.

\vspace*{-2ex}
\section*{\sc acknowledgement}
\vspace*{-1ex}
\indent \indent Authors thank to Science Data Center of JHU/APL for AMPTE/CCE FGM dataset, V.~Angelopoulos and THEMIS team for providing THEMIS FGM and ESA datasets, CDAWeb for providing Cluster, Polar, and Wind datasets which are shown in the paper.

\end{multicols}

\end{document}